# Spectroscopic studies of the two EROS candidate microlensed stars. *

J.P.Beaulieu[1], R.Ferlet[1], P.Grison[1], A.Vidal-Madjar[1], J.P.Kneib[2], E.Maurice[3], L.Prévot[3], C.Gry[4], J.Guibert[5], O.Moreau[5], F.Tajhmady[5], E.Aubourg[6], P.Bareyre[6], C.Coutures[6], M.Gros[6], B.Laurent[6], M.Lachièze-Rey[6], E.Lesquoy[6], C.Magneville[6], A.Milsztajn[6], L.Moscoso[6], F.Queinnec[6], C.Renault[6], J.Rich[6], M.Spiro[6], L.Vigroux[6], S.Zylberajch[6], R.Ansari[7], F.Cavalier[7], and M.Moniez[7]

[1] Institut d'Astrophysique de Paris, CNRS, 98bis boulevard Arago 74014 Paris France.
[2] ESO La Silla casilla 19001 Santiago 19, Chile.
[3] Observatoire de Marseille, 2 place Le Verrier, 13248 Marseille 04, France.
[4] Laboratoire d'Astronomie Spatiale CNRS, Traversée du siphon, les Trois Lucs, 13120 Marseille, France.
[5] Centre d'Analyse des Images de l'Institut National des Sciences de l'Univers, CNRS Observatoire de Paris, 61 Avenue de l'Observatoire, 75014 Paris, France.
[6] CEA, DSM/DAPNIA, Centre d'études de Saclay, 91191 Gif-sur-Yvette, France.
[7] Laboratoire de l'Accélérateur Linéaire IN2P3, Centre d'Orsay, 91405 Orsay, France.

**Abstract.** Low resolution spectroscopy, and UBVRI photometry, have been obtained for the two EROS microlensing candidates. Radial velocities indicate that both stars are members of the Large Magellanic Cloud. The spectrum and the colours of EROS1, the first candidate, reveal that it is a moderately reddened main-sequence B star with H emission lines. The presence of H absorption lines seems to be more the signature of a normal star than that of a cataclysmic variable. As to EROS2, the second candidate, its spectrum and photometry are those of an unreddened normal main-sequence A star, but it cannot be totally excluded that they represent those of a nova in the pre-outburst phase. Although it is not yet possible to exclude intrinsic stellar variations, the interpretation in terms of microlensing effects remain the most natural one.

**Key words:** : Stars : fundamental parameter - Stars : emission-line, Be - Cosmology : dark matter - Cosmology : gravitational lensing

## 1. Introduction

The EROS (Expérience de Recherche d'Objets Sombres) French collaboration (Aubourg et al., 1993a, 1993b), and the American-Australian MACHO project (Alcock et al.,1993) are both searching for baryonic dark matter in the galactic halo through microlensing effects on stars of the Large Magellanic Cloud (Paczyński, 1986). The OGLE project (Udalski et al., 1993) is searching for microlensing effects due to disk objects on stars in the bulge of our galaxy.

Due to the transverse motion of the galactic halo with respect to the LMC-observer direction, a dark object of the galactic halo (a brown dwarf or, with lower probability, a white dwarf, a neutron star or a black hole) passing by closely enough to the line of sight of a background star in the LMC induces an increase on the apparent brightness of the star. The observational characteristics of such an event are:

-The brightening amplification, larger than 0.3 magnitude if the deflector distance to line-of-sight is smaller than the Einstein radius. This corresponds to a probability of $0.5 \; 10^{-6}$ of finding any given star amplified by 1.34 at a given time. A photometric precision of the order of 0.1 mag is needed to measure such an amplification.

-The characteristic time scale of the event, roughly centered around 0.03 to 30 days when assuming a standard isothermal halo and $10^{-7} \; M_\odot < M < 10^{-1} \; M_\odot$, M being the mass of the deflector.

-The light curve, symmetric in time, achromatic and transient.

Between 1990 and 1993, more than 150 B and R plates have been obtained at the ESO 1m Schmidt telescope. After digitization on the MAMA (Machine Automatique à Mesurer pour l'Astronomie, Berger 1991) at the Observatoire de Paris, the data have been processed by a specially-devised program of image reconstruction and stellar photometry (Aubourg et al., 1993b, Aubourg et al., 1994 en preparation). About four million stars have been passed through a filter designed to select the and EROS2), and only two, showing photometric variations compatible with the characteristics of microlensing events have been found by the EROS collaboration and their photometric curves have been published by Aubourg et al., 1993b. Their positions are :

EROS1 : $\alpha$=5h 26m 34s   $\delta$ = -70° 57′ 45″ (2000.0)
EROS2 : $\alpha$=5h 06m 05s   $\delta$ = -65° 58′ 34″ (2000.0)

The aim of this paper is to present additional spectroscopic and photometric observations that we have performed to derive the physical properties of the microlensed stars in order to have additional evidence whether we are dealing with true microlensing events, or with any kind of known-type variable, or with a new, extremely-rare intrinsic phenomenon.

## 2. Observations

### 2.1. Photometry

The UBVRI observations were carried out with the 1.54m Danish telescope at La Silla. The UBVR filters are those of Bessel, I is the Gunn's filter. The measurements were corrected for atmospheric extinction by using the standard coefficients at La Silla (ESO Users manual 1993); the primary standard stars were selected in the E4 region published by Graham (1982). A secondary standard sequence, VA6, (Vigneau and Azzopardi, 1982) was also observed. Data reduction was achieved by means of an aperture photometry routine performed in the MIDAS environment. The log of observations is given in Table 1; V magnitudes and colours indices of both candidates are gathered in Table 2. The uncertainty given in this Table is simply the average of the differences between magnitudes derived from the two standard sequences.

**Table 1.** Log of the photometric observations

| Field centre | Date | Filter | Exp.time | Seeing |
|---|---|---|---|---|
| EROS1 | 94-01-14 | U | 30min | 1" |
| EROS1 | 94-01-14 | B | 15min | 1" |
| EROS1 | 94-01-14 | V | 10min | 1" |
| EROS1 | 94-01-14 | R | 10min | 1" |
| EROS1 | 94-01-18 | I | 10min | 1.3" |
| EROS2 | 94-01-17 | U | 20min | 1.3" |
| EROS2 | 94-01-17 | B | 15min | 1.3" |
| EROS2 | 94-01-17 | V | 10min | 1.3" |
| EROS2 | 94-01-17 | R | 10min | 1.3" |
| EROS2 | 94-01-17 | I | 10min | 1.3" |

### 2.2. Spectroscopy

The spectroscopic observations were performed at the ESO 3.5m NTT telescope, equipped with EMMI, in the red medium dispersion mode. The detector was a LORAL UV coated chip of 2048 x 2048 pixels, the nominal gain is 1.5 e-/ADU, and the read out noise 7 e-. The projected width of the 1" entrance slit is about 3 pixels on the detector, i.e. 500 km.s$^{-1}$ : Table 3 gives the log of observations.

**Table 2.** UBVRI photometry of EROS1 and EROS2

| Star | V | U-B | B-V | V-R | R-I | errors |
|---|---|---|---|---|---|---|
| EROS1 | +19.11 | -0.22 | +0.35 | +0.39 | +0.21 | ±0.1 |
| EROS2 | +19.38 | -0.02 | +0.04 | +0.09 | +0.13 | ±0.1 |

**Table 3.** Log of the spectroscopic observations. The estimation of the signal-to-noise ratio has been made at 600 nm.

| Date | star | resolution | S/N |
|---|---|---|---|
| 93-10-18 | EROS1 | 1.48 nm | 10 |
| 93-10-19 | EROS2 | 1.48 nm | 10 |
| 94-2-1 | EROS2 | 1.11 nm | 15 |

The spectroscopic observations were reduced at the Institut d'Astrophysique de Paris using the MIDAS X spectra context and the standard procedures of reduction. Bias subtraction, flat-fielding, cosmic rays removal and extinction corrections were performed in a standard way. Extraction of one-dimensional spectra was optimized using algorithms derived from the weighted average methods developed by Horne (1986) and by Robertson (1986). The EMMI instrumental response was corrected by observing the standard star GD50 taken from the list of Oke (1990). The spectra obtained were smoothed through a median filter and normalized to unity. The absolute wavelength calibration was achieved using the sky emission lines of O I (636.40 nm, 630.13 nm, 557.81 nm), and N I (520.17 nm, 522.89 nm). We estimate that the precision of the zero-point determination and the uncertainty in line positions are better than 0.1nm. These values do not allow a derivation of accurate stellar velocities, but at least they enable to establish the cloud membership or the galactic nature of the microlensed star candidates.

**Table 4.** Table 4 provides line identifications and equivalent widths determined from the three usable spectra. Their low resolution prevents us from assigning a refined spectral type. The symbol 'd' means that the line is detected but its equivalent width is not measurable.

| Line | E.W.(nm) EROS1 | E.W.(nm) EROS2 |
| --- | --- | --- |
| H$\alpha$ | 0.7 $\pm$ 0.1 (emission) | 1.3 $\pm$ 0.1 |
| H$\beta$ | d. (emission) | 1.4 $\pm$ 0.1 |
| H$\gamma$ | 1.2 $\pm$ 0.1 | 1.6 $\pm$ 0.1 |
| H$\delta$ | d | 1.5 $\pm$ 0.1 |
| CaII H+H $\epsilon$ | d | d |
| CaII K | | d |
| Na D1D2 | | d |

The radial velocity of EROS1, estimated from the emission line of H$\alpha$ on the spectrum of Oct. 18 1993, is 350 $\pm$ 170 km.s$^{-1}$. On the single velocity criterion, EROS1 is obviously a member of the LMC, but because of the large uncertainty in this determination, it is not possible to go into more details as to its agreement with the local LMC velocity field. With the classical LMC distance modulus of 18.5, EROS1 has an absolute magnitude of -0.7, corresponding to a late B-type star. Its spectrum (Figure 1 and Table 4) is undoubtedly that of a Be star as it shows the typical H$\alpha$ emission together with the H$\beta$ line filled by emission. The H$\alpha$ width at the base of the line is about 1500 km s$^{-1}$, quite comparable with those of many classical Be stars. The line shows no central reversal. As helium lines, on which the classification of normal stars is based, are not detected at such a dispersion, MK classification cannot be performed and our attempt to assign a spectral type to this star rests on the calibration of MK spectral types as a function of equivalent widths of Balmer lines. This work has been carried out by Didelon (1982) for H$\alpha$, H$\beta$ and various lines of other elements. Didelon's calibration indicates a spectral type about A0III-V which, based on a single line, is considerably uncertain but consistent with the Be nature of EROS1. The observed colours of EROS1 are on average in agreement with those of some Be stars in the lists published by Schild (1978). Like for all Be stars, the flux of EROS1 is difficult to derive from its observed colours with the present set of data because there is no way to determine accurately the respective amount of circumstellar and interstellar reddening. It is even more difficult in the present case as our photometry is, on average, uncertain by about 0.1 mag. However, results obtained by Schild (1978) show that the intrinsic reddening E(B-V) of the Be stars in the visible range follows more or less the interstellar reddening law in $\exp(-1/\lambda)$. Dereddening the observed colours by E(B-V)=0.42 by adopting the extinction coefficients determined by Savage and Mathis (1979), provides the following colour indices:

(U-B)$_0$=-0.49 (B-V)$_0$=-0.07 (V-R)$_0$=+0.06 (R-I)$_0$=-0.13
Which, on average correspond to the intrinsic colours of a B8 excess.

This result was checked by using the Q-method introduced by Johnson and Morgan (1953) and rediscussed by Gutiérez-Moreno (1975). That the Q-method may be validly applied to Be stars is a matter of questioning because of the dual nature of its reddening. It is clear that, it must provide too early spectral type because the intrinsic reddening is considered as interstellar reddening. Let us recall that the calibration of
$Q = (U - B) - \frac{E_{U-B}}{E_{B-V}}(B - V)$
allows to determine the spectral type, the interstellar reddening and the intrinsic colours of the B0-A0 main sequence stars. Here we get Q=0.45. Using the Gutiérez-Moreno (1975) calibration, we derive a spectral type B6 from which we get a total reddening $E_{B-V} = 0.49$. Hence we find as intrinsic colours

(U-B)$_0$=-0.55 (B-V)$_0$=-0.17 (V-R)$_0$=-0.01 (R-I)$_0$=-0.22
The spectral type and the colours obtained by this method are significantly bluer than the ones obtained above. This discrepancy is induced by the use of the spectral type, $(B-V)_0$ relation which provides a B-V colour excess larger than the one obtained through deredenning with Savage and Mathis extinction coefficients.

EROS1 has an absolute magnitude of -0.7, corresponding to a B6-7 IVe or B6-7 Ve after the calibrations provided by Schmidt-Kaler (1982) and Zorec and Briot (1993).

Finally, it must also be noticed that CCD images show that EROS1 most probably has a close companion, the centers of both components being separated by a small fraction of the FWHM. We estimated the EROS1 companion to be at the same right ascension and $-1.1" \pm 0.3$ in declination. It is about 2.6 $\pm$ 0.5 magnitudes fainter than EROS1. The intrinsic colors of this star could not be derived.

The radial velocity of EROS2, based on the four visible Balmer lines, is 300 $\pm$ 150 km.s$^{-1}$, a value which confirms its membership to the LMC like for EROS1. The spectrum of EROS2 suffers from a limited resolution. In the same way we have used Didelon's calibration supplemented by the equivalent widths of Balmer lines H$\alpha$ to H$\delta$ measured photo-electrically by Tereshchenko (1976). The derived spectral type B9III-V should be accurate within 2 or 3 subclasses. UBVRI colours are those of an unreddened (or very slightly reddened) main sequence A0-2 star; the luminosity class V results from its absolute magnitude equal to 0.9. These results are quite consistent with the spectral type derived from the measurements of equivalent widths.

Similarly to EROS1, EROS2 is suspected to have a close companion from inspection of its profiles on multi-colour CCD frames. It would be at the following position, $-0.1" \pm 0.2"$ in right ascension, and $-0.7" \pm 0.2"$ in declination. It is about 3 $\pm$ 0.5 magnitudes fainter than EROS2, and its intrinsic colors could not be derived.

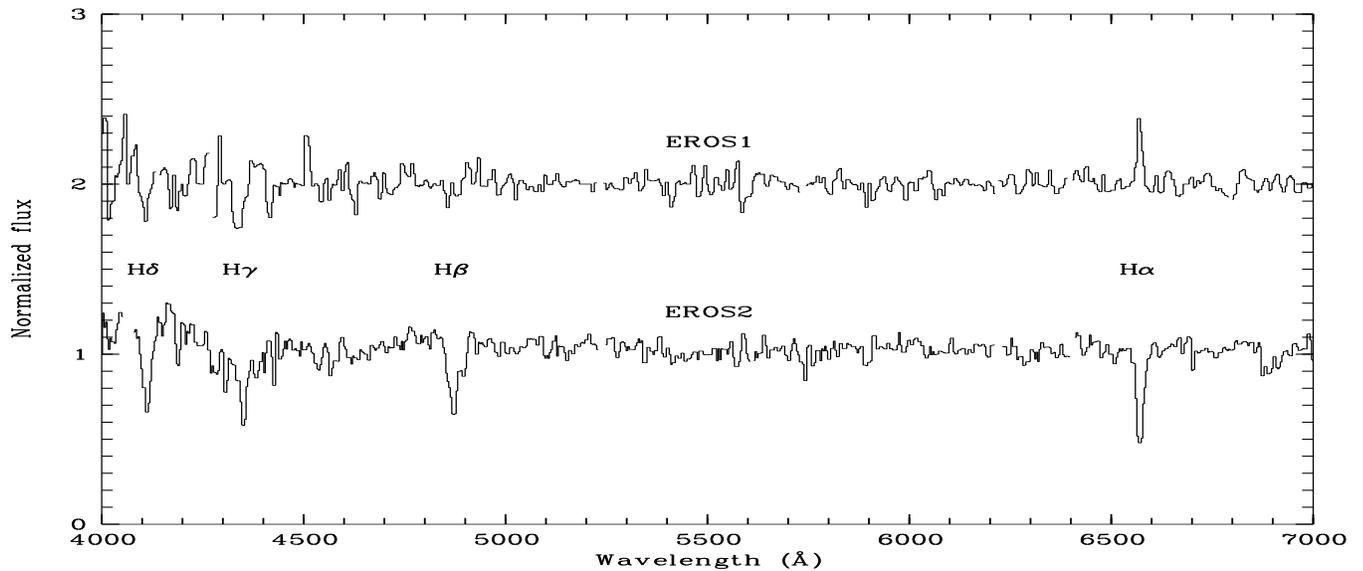

**Fig. 1.** Figure 1 shows the spectra of EROS1 and EROS2 in their most significant region. Cosmic rays have been removed. Parts of spectra containing artifacts like dark spots and/or cold columns have been suppressed, except the feature at 450.8nm.which could result from an unfiltered cosmic ray.

## 4. Discussion

One characteristics of a microlensing event must be its random nature.

Spectroscopically, EROS2 is an A0-A2 main sequence star, i.e. a very common object in the sample we have monitored. Oppositely, the peculiarity of EROS1, which is undoubtedly a Be star, requires to discuss the abundance of such objects in the LMC. The proportion of Be among galactic B stars has been analyzed by Briot and Zorec (1993) from a sample of stars brighter than V=7.1. Their figure 1 shows that only 5 percent among the B6-7 stars are expected to be Be stars. Assuming that this ratio is valid for the LMC makes these stars rather uncommon objects and may eventually lead to question the reality of the microlensing event.

Photometrically, since the beginning of the search for dark matter in the galactic halo via the EROS and MACHO experiments, investigations are made to determine the types of variables which could produce light curves similar to those resulting from microlensing events as calculated by Paczyǹski (1986).To our knowledge, such theoretical curves, characterized by the time-symmetry, the achromaticity and the transience of the brightening, cannot be adequately represented by any type of known variables. However, the limited accuracy of experimental microlensing light curves makes possible their interpretation by a very small number of variables from which Be stars must be excluded because, up to now, none has provided similar variations. Some stars, like HD 120991 (Waelkens et al., 1983) have shown an increase of brightness suggesting a microlensing event from its time scale and symmetry but neither from the unicity nor the achromaticity criteria. Furthermore, the V fluctuation of HD 120991 was about 0.2 magnitude, much less than the one observed in EROS1.

The objects which could give an acceptable, non-microlensing explanation of the photometric variations of EROS1 and EROS2 belong to the class of eruptive- cataclysmic-variables:

1) Old novae like GK Per show among their apparently erratic outbursts some flares which have time scales, time symmetry and amplitudes quite similar to experimental gravitational microlensing curves. A striking example is provided by the 1973, 1975 or 1981 bursts visible on the visual light curve of GK Per published by Sabadin and Bianchini (1983). The achromaticity of these outbursts remains however uncertain because of the scarcity of B measurements performed during the events. These objects are usually the hot components of binary systems consisting of an under-luminous blue star and a late-type companion. Their spectra present a considerable variety (Kraft, 1964), most of them showing emission lines of H, often of late-type being visible in some cases. The ratio of the H$\alpha$ and H$\beta$ emissions in the spectrum of EROS 1 is comparable to that of variables like RS Oph (Williams, 1983) or some other recurrent novae. But as no evidence of absorption lines has been found in the spectra of past novae after Hack et al.(1993) and as no late-type companion is visible on the spectrum of EROS1, we are inclined to conclude that the blue star has probably a normal luminosity and, consequently, that it is unlikely that EROS1 may be a recurrent nova. But obviously long term and more detailed observations must be collected before being able to give a conclusive answer to this question.

2) Pre novae, some light curves of which have been published by Robinson (1975). Nova Sgr 1962 is an example of the similarity between microlensing events and pre-eruption light curves of some novae. Spectra of three stars only (V603 Aql, V533 Her and HR Del) have been observed in the phase preceding their outburst. According to Hack et al. (1993), the energy distribution of V603 Aql is similar to that of B and A stars and the Balmer series is detectable in absorption at low resolution, while the energy distributions of V533 Her and HR Del resemble those of O or early B stars without any emission or absorption. Therefore the B9-A2 type we assigned to EROS2 makes it marginally compatible with the category of pre-novae. Similarly to EROS1, more observations of this star are needed before being able to know its true nature.The absolute magnitudes of EROS1 and EROS2 are at the brighter limit of slow novae in quiescent phases, after Figure 2 and Table 1 of Della Valle and Duerbeck (1993). However, this argument cannot be used to reject the slow-nova interpretation when considering the possibility of both candidates to be double stars.

Finally, our conclusions are quite similar to those derived by Della Valle (1994) for the MACHO candidate: even if the observations are marginally compatible with the old-pre novae interpretation just as the possibility of a class of variable stars yet unknown has to be considered, the gravitational lensing yet remains a more satisfactory explanation, but a final proof of this assumption remains to be achieved by means of extended observations. In particular, high resolution images of the EROS and MACHO candidates should greatly help to clarify the question of their environment. Accurate long time-base photometry is also in progress.

*Acknowledgements.* We thank C. Bellanger, V. de Lapparent and Y. Mellier for having kindly given us some of their observing time, M. Della Valle for doing some observations, P. Leisy, J. Zorec, M. Gerbaldi for useful discussions. We are grateful for the support given to our project by the technical staff at ESO La Silla.